\documentclass[12pt]{wlscirep}
\usepackage[utf8]{inputenc}
\usepackage[T1]{fontenc}
\usepackage{hyperref}
\usepackage[final]{pdfpages}
\makeatletter
\AtBeginDocument{\let\LS@rot\@undefined}
\makeatother
\hypersetup{
    colorlinks = false,
    linkbordercolor = {white},
}

\title{Nearest neighbor permutation entropy detects phase transitions in complex high-pressure systems}

\author[1,*]{\small Arthur A. B. Pessa}
\author[1]{\small Leonardo G. J. M. Voltarelli} 
\author[2,3,4]{\small L\'ucio Cardozo-Filho} 
\author[2]{\small Andres G. M. Tamara} 
\author[5]{\small Claudio Dariva} 
\author[3]{\small Papa M. Ndiaye} 
\author[6,7,8,9,10,$\dagger$]{\small Matja{\v z} Perc} 
\author[1,$\ddagger$]{\small Haroldo V. Ribeiro} 

\affil[1]{\footnotesize Departamento de F\'isica, Universidade Estadual de Maring\'a - Maring\'a, PR 87020-900, Brazil}

\affil[2]{\footnotesize School of Engineering, Sao Paulo State University (UNESP), Campus of Sao Joao da Boa Vista, Sao Joao da Boa Vista, SP, 13876-750, Brazil}
\affil[3]{\footnotesize Programa de Engenharia Qu\'imica / COPPE, Universidade Federal do Rio de Janeiro, Cidade Universit\'aria, CP: 68502, Rio de Janeiro, RJ 21941-972, Brazil}
\affil[4]{\footnotesize Graduate Program in Chemical Engineering, State University of Maringa, Av. Colombo 5790, Maringa, Brazil}

\affil[5]{\footnotesize Center for Studies on Colloidal Systems (NUESC), Institute of Technology and Research (ITP), Av. Murilo Dantas, 300, Farol\^andia, Aracaju, SE, Brazil}

\affil[6]{\footnotesize Faculty of Natural Sciences and Mathematics, University of Maribor, Koro{\v s}ka cesta 160, 2000 Maribor, Slovenia}
\affil[7]{\footnotesize Community Healthcare Center Dr. Adolf Drolc Maribor, Ulica talcev 9, 2000 Maribor, Slovenia}
\affil[8]{\footnotesize Department of Physics, Kyung Hee University, 26 Kyungheedae-ro, Dongdaemun-gu, Seoul 02447, Republic of Korea}
\affil[9]{\footnotesize Complexity Science Hub, Metternichgasse 8, 1030 Vienna, Austria}
\affil[10]{\footnotesize University College, Korea University, 145 Anam-ro, Seongbuk-gu, Seoul 02841, Republic of Korea}

\affil[*]{\footnotesize email: aabpessa@uem.br}
\affil[$\ddagger$]{\footnotesize email: matjaz.perc@gmail.com}
\affil[$\ddagger$]{\footnotesize email: hvr@dfi.uem.br}

\begin{abstract}
Understanding the high-pressure phase behavior of carbon dioxide–hydrocarbon mixtures is of considerable interest owing to their wide range of applications. Under certain conditions, these systems are not amenable to direct visual monitoring, and experimentalists often rely on spectrophotometric data to infer phase behavior. Consequently, developing computationally efficient and robust methods to leverage such data is crucial. Here, we combine nearest neighbor permutation entropy, computed directly from \textit{in situ} near-infrared absorbance spectra acquired during depressurization trials of mixtures of carbon dioxide and a distilled petroleum fraction, with an anomaly detection approach to identify phase transitions. We show that changes in nearest neighbor entropy effectively signal transitions from initially homogeneous mixtures to two-phase equilibria, thereby enabling accurate out-of-sample online predictions of transition pressures. Our approach requires minimum data preprocessing, no specialized detection techniques or visual inspection of the spectra, and is sufficiently general to be adapted for studying phase behavior in other high-pressure systems monitored via spectrophotometry.
\end{abstract}

\begin{document}
\rfoot{\small\sffamily\bfseries\thepage/15}%
\flushbottom
\maketitle
\thispagestyle{empty}

\section*{Introduction}\label{sec:intro}

Carbon dioxide flooding is widely employed as a miscible displacement technique to enhance oil recovery through mechanisms such as crude vaporization, oil swelling, viscosity reduction, and interfacial tension modification~\cite{almobarak2021areview, li2019molecular}. Furthermore, this technology facilitates the underground storage of substantial quantities of ${\rm CO}_2$, thereby contributing to carbon footprint reduction~\cite{rodrigues2022multi-objective, zhao2023amulti-medium}. Conversely, many pre-salt reservoirs along the Brazilian coast contain high concentrations of carbon dioxide, rendering high-pressure phase-equilibrium data essential for petroleum reservoir exploration and planning. Moreover, mixtures of ${\rm CO}_2$ and hydrocarbons typically exhibit complex phenomena, including liquid-liquid and vapor-liquid-liquid equilibria, and in some cases, homogeneous mixtures cannot be achieved even at very high pressures~\cite{yanes2020study, borges2015near, cruz2019co2influence}.

Traditional approaches to understanding these phase behaviors have typically involved thermodynamic analyses of variations in pressure, volume, and temperature~\cite{alher2018estimating, regueira2020density, lucas2016useofreal}. However, the direct visualization methods employed in these approaches are limited by the opacity of certain crude oils. Studies indicate that most crude oils exhibit a minimum absorbance near $1.6$ $\mu{\rm m}$, rendering visualization techniques operating in the visible spectrum inefficient for some assessments~\cite{daridon2020combined, yanes2021phase}. Despite the extensive body of work on the phase behavior of hydrocarbon mixtures, there remains a need to develop new tools and strategies to probe the complex physicochemical phenomena involved.

In this context, the application of spectroscopy to monitor phase equilibrium in complex chemical systems at high pressures is an important subset of analytical methods that minimizes perturbations during sample withdrawal~\cite{dohrn2012experimental, braeuer2015situ, dohrn2024high-pressure}. Furthermore, combining spectroscopic techniques with robust, computationally efficient information‐theoretic methods -- such as permutation entropy~\cite{bandt2002permutation} and related ordinal approaches~\cite{zanin2012permutation, riedl2013practical, amigo2015ordinal, keller2017permutation, pessa2021ordpy} -- has emerged as a promising strategy for investigating phase behavior under high-pressure conditions, as evidenced by its successful application in diverse spectroscopic and phase behavior studies~\cite{li2017application, li2017pretreatment, garland2018anomaly, ren2021rescaled, zhang2022detecting, du2023interconnected, du2020detecting, sigaki2019estimating, pessa2021determining}. For instance, permutation entropy has been employed to segment near-infrared spectra from blood samples, thereby identifying wavelength ranges that improve blood glucose prediction~\cite{li2017application, li2017pretreatment}. It has also served as an anomaly detection tool in time series of relative isotope abundances derived from laser absorption spectroscopy of ice cores~\cite{garland2018anomaly}. Ordinal methods have been extensively applied to characterize distinct flow regimes -- such as slug flow, non-uniform bubble flow, and uniform bubble flow -- in two-phase and three-phase systems~\cite{ren2021rescaled, zhang2022detecting, du2020detecting, du2023interconnected}. Moreover, permutation entropy and ordinal methods have yielded intriguing insights into phase transitions in thermotropic liquid crystals~\cite{sigaki2019estimating, pessa2021determining}.

Here, we present an application of the recently proposed $k$-nearest neighbor permutation entropy~\cite{voltarelli2024characterizing} to experimental transflectance spectra obtained \textit{in situ} from mixtures of carbon dioxide and a distilled petroleum fraction under controlled high-pressure conditions. Specifically, we develop a methodology to detect phase transitions via abrupt changes in the entropy of the measured spectra. Our approach operates online, detecting phase transitions in real time as new sample spectra are acquired. Furthermore, because the $k$-nn permutation entropy -- as well as other ordinal methods~\cite{bandt2002permutation, pessa2021ordpy} -- relies on the relative amplitudes of neighboring data points, our method requires minimum data preprocessing, no specialized detection techniques, and no visual inspection of the spectra, and it can be readily extended to other systems monitored via spectrophotometric measurements.

The remainder of this paper is structured as follows. First, we present the dataset of absorbance spectra collected during depressurization trials of mixtures of carbon dioxide and hydrocarbon fractions. Next, we detail the formalisms underlying both the $k$-nearest neighbor permutation entropy and permutation entropy before their applications to the spectra. We then describe our anomaly detection approach for inferring phase transitions and present the results of its application. Finally, we discuss our findings and conclude.

\section*{Data}\label{sec:data}

\begin{figure*}[!ht]
  \centering
  \includegraphics[width=1.00\textwidth, keepaspectratio]{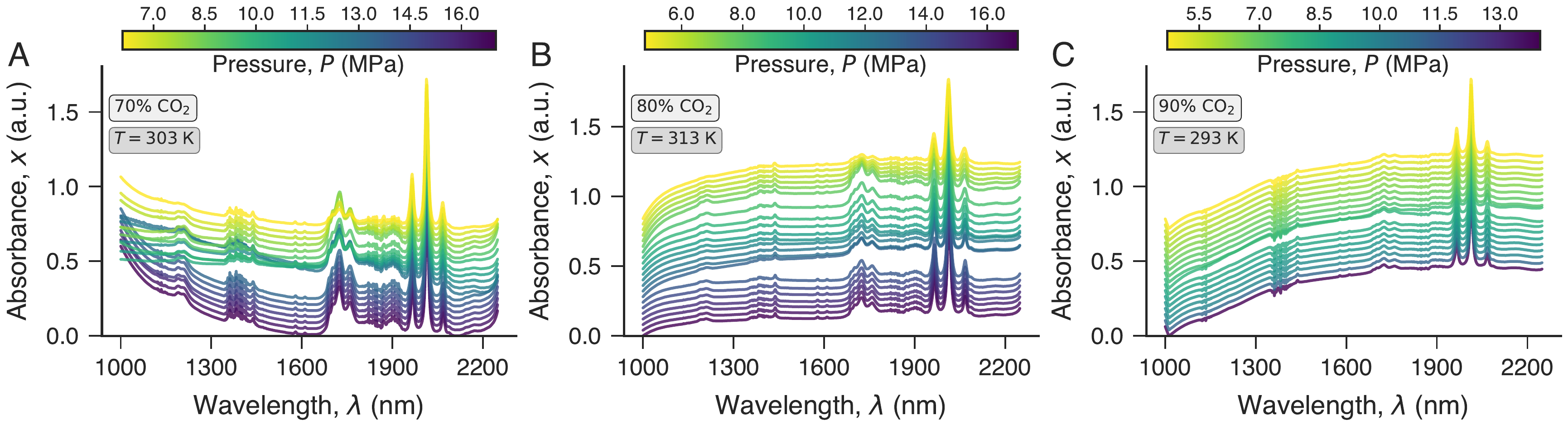}
  \caption{Near-infrared spectra of mixtures of carbon dioxide and distilled petroleum fractions. Panels (A), (B), and (C) show absorbance spectra (in arbitrary units) for wavelengths between $1000$ nm and $2250$ nm measured at various pressures for three distinct samples. The spectra are color-coded by pressure (see colorbars) and annotated with the relative ${\rm CO}_2$ sample weight and temperature. Absorbance spectra for each depressurization trial were recorded at $0.1$ MPa increments and shifted in these visualizations for clarity.
}
\label{fig:1}
\end{figure*}

The dataset used in our investigations was originally introduced by Borges \textit{et al.}~\cite{borges2015near} and comprises near-infrared (NIR) absorbance spectra acquired via transflectance measurements during depressurization trials of mixtures of carbon dioxide (${\rm CO}_2$) and distilled petroleum fractions maintained within a high‐pressure and variable‐volume view cell. The distilled fraction is light yellow in color, with a density of $0.8230~\text{g}\,\text{cm}^{-3}$ at 298~K and a normal boiling point of 569~K; the raw oil composition, determined through standard fractionation analysis, includes 74.7\% saturates, 14.7\% aromatics, 10.6\% resins, and less than 0.5\% asphaltenes by weight~\cite{borges2015near}. The spectra, denoted by $\{x_\lambda\}_{\lambda = \lambda_{\rm min}, \dots, \lambda_{\rm max}}$, represent the sample absorbance ($x_\lambda$) indexed by 1440 distinct wavelengths ($\lambda$) unevenly sampled between $\lambda_{\rm min} = 1000$ nm and $\lambda_{\rm max} = 2250$ nm (Figure~\ref{fig:1}). Mixtures were prepared using six different ${\rm CO}_2$ concentrations, corresponding to relative sample weights ranging from $35\%$ to $90\%$. Subsequently, several depressurization runs were performed at fixed temperatures ($T = 293$ K, $T = 303$ K, $T = 313$ K, and $T = 333$ K), while the samples were subjected to pressures ranging from $4.1$ MPa to $19.5$ MPa (see Table~\ref{tab:1} for an overview of pressure intervals and the number of runs performed at different ${\rm CO}_2$ concentrations and temperatures). Each trial yielded between $21$ and $127$ spectra, with measurements recorded at successive pressure increments of $0.1$ MPa. For each combination of ${\rm CO}_2$ concentration and temperature, the transition pressure $P_{t}^{\rm exp}$ corresponding to the observed phase transitions during depressurization was visually determined by an expert. A more detailed description of the experimental apparatus and techniques used in data collection is provided in Ref.~\cite{borges2015near}. 

The observed phase transitions encompass transitions from homogeneous liquid or gas mixtures (at higher pressures) to different types of two-phase equilibria (at lower pressures), as well as a transition from a two-phase liquid mixture to a multiphase system (at even lower pressures). We categorize these transitions as follows: 
\begin{itemize}
    \item liquid-liquid equilibrium (LLE): transitions from a homogeneous liquid mixture to a two-phase liquid mixture;
    \item vapor-liquid equilibrium via bubble point (VLE-BP): transitions from a homogeneous gas mixture to a two-phase gas-liquid mixture upon crossing the bubble-point curve;
    \item vapor-liquid equilibrium via dew point (VLE-DP): transitions from a homogeneous liquid mixture to a two-phase gas-liquid mixture upon crossing the dew-point curve; 
    \item vapor-liquid-liquid equilibrium (VLLE): transitions from a two-phase liquid mixture to a multiphase system with three coexisting phases at equilibrium (one gas phase and two immiscible liquids).
\end{itemize}
Supplementary Table~S1 summarizes the key variables extracted from this dataset.

\begin{table}[!ht]
    \centering
    \caption{Summary of the carbon dioxide and distilled petroleum fraction mixtures used in depressurization trials. Six samples were prepared at six different carbon dioxide (${\rm CO}_2$) concentrations relative to sample weight, with depressurizations conducted at four different temperatures. The table shows the number of depressurization runs for each sample as well as the corresponding pressure intervals.
    }\label{tab:1}
    \begin{small}
    \begin{tabular}{c|c|c|r}
    \hline \hline
    ${\rm CO}_2$ &Temperature&Runs& Pressure range  \\\hline \hline
    35\%         &$293$ K    & 4  & 5.3--10.0 MPa   \\
    35\%         &$303$ K    & 3  & 5.4--12.0 MPa   \\
    35\%         &$313$ K    & 2  & 8.0--13.5 MPa  \\
    35\%         &$333$ K    & 2  & 11.0--16.0 MPa  \\[.35em]
    45\%         &$293$ K    & 3  & 4.9--13.0 MPa   \\
    45\%         &$313$ K    & 2  & 7.2--16.0 MPa   \\
    45\%         &$333$ K    & 2  & 12.6--17.5 MPa  \\[.35em]
    60\%         &$303$ K    & 2  & 6.1--17.0 MPa   \\
    60\%         &$313$ K    & 2  & 12.8--16.0 MPa  \\
    60\%         &$333$ K    & 2  & 16.0--18.5 MPa  \\[.35em]
    70\%         &$303$ K    & 2  & 5.8--17.5 MPa   \\
    70\%         &$313$ K    & 2  & 13.5--17.5 MPa  \\
    70\%         &$333$ K    & 2  & 16.4--19.5 MPa  \\[.35em]
    80\%         &$313$ K    & 2  & 4.8--17.0 MPa   \\
    80\%         &$333$ K    & 2  & 15.5--18.5 MPa  \\[.35em]
    90\%         &$293$ K    & 2  & 4.1--14.0 MPa   \\
    90\%         &$303$ K    & 2  & 6.2--12.0 MPa   \\
    90\%         &$313$ K    & 2  & 10.5--13.0 MPa  \\
    90\%         &$333$ K    & 2  & 15.0--18.0 MPa  \\
    \hline \hline
    \end{tabular}
    \end{small}
\end{table}

\section*{Methods}\label{sec:methods}

To investigate whether phase transitions can be detected through changes in the characteristics of sample spectra during depressurization runs, we use the $k$-nearest neighbor ($k$-nn) permutation entropy~\cite{voltarelli2024characterizing}. This information-theoretic measure arises from combining the permutation entropy framework~\cite{bandt2002permutation, zanin2012permutation, riedl2013practical, amigo2015ordinal, keller2017permutation, pessa2021ordpy} -- originally developed for time series analysis -- with biased random walks such as those implemented in node2vec~\cite{grover2016node2vec}, an algorithm for obtaining high-dimensional vector representations of vertices in complex networks. The $k$-nn permutation entropy has been primarily proposed to address unstructured, irregularly sampled datasets such as our spectra, and its application to structured data (time series and images) has also yielded excellent results~\cite{voltarelli2024characterizing}.

\begin{figure*}[ht!]
\centering
\includegraphics[width=0.92\textwidth]{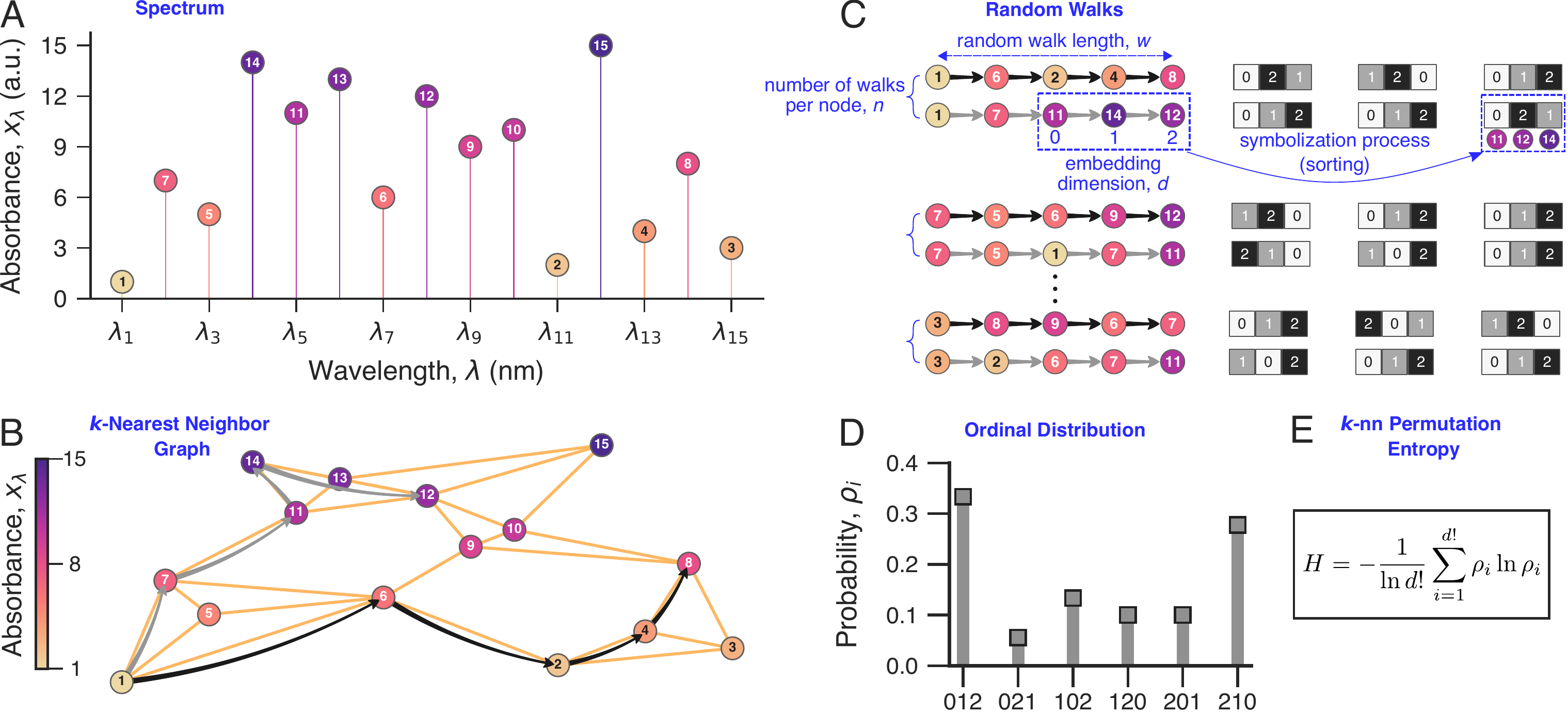}
\caption{Calculation of the $k$-nearest neighbor permutation entropy for sequential data. (A) Illustration of a short, unevenly sampled, hypothetical spectrum $\{x_\lambda\}_{\lambda = \lambda_1, \dots, \lambda_{15}}$. (B) From this hypothetical spectrum, we construct a $k$-nearest neighbor graph ($k=3$ in this example) using the data coordinates $(\lambda, x_\lambda)$ to define neighborhood relationships. In this graph, each observed absorbance value $x_\lambda$ is represented by a node, with undirected edges connecting pairs of observations $x_{\lambda_i} \leftrightarrow x_{\lambda_j} $ when $(\lambda_j, x_{\lambda_j})$ is among the $k$-nearest neighbors of $(\lambda_i, x_{\lambda_i})$. For ease of interpretation, the min–max normalization step has been omitted from this illustration. (C) Subsequently, we execute $n$ biased random walks of length $w$ starting from each node, sampling the absorbance values $(x_\lambda)$ to generate time series ($n=2$ and $w=5$ in this example). We then apply the Bandt-Pompe symbolization approach to each of these time series. This symbolization entails creating overlapping partitions of length $d$ ($d=3$ in this example) and arranging partition indices in ascending order to determine the sorting permutation for each partition. (D) Finally, we evaluate the probability of occurrence of each of the $d!$ permutation types (ordinal distribution) and calculate its (E) Shannon entropy, thereby defining the $k$-nearest neighbor permutation entropy.}
\label{fig:2}
\end{figure*}

The initial step in calculating the $k$-nn entropy involves fixing the number of neighbors $k$ to construct a nearest neighbor graph on which random walks are performed (Figures~\ref{fig:2}{A} and~\ref{fig:2}{B}). In our dataset, each absorbance value $x_\lambda$ of a spectrum $\{x_{\lambda}\}_{\lambda = \lambda_{\rm min}, \dots, \lambda_{\rm max}}$ is mapped to a node in the graph. Both the absorbance and its corresponding wavelength $(\lambda, x_{\lambda})$ are considered when defining the nearest neighbors via the Euclidean distances between all datapoints $(\lambda, x_{\lambda})$ in the spectrum. However, the range of variation and the typical scale of differences between wavelengths and absorbances are not equivalent. In fact, the locally smooth behavior of the spectrum renders neighboring data points considerably closer in terms of absorbance than in terms of wavelength, thereby disproportionately weighting the absorbance values during graph construction. To mitigate this imbalance, we rescale the wavelength and absorbance values to the interval $[0,1]$ via min-max normalization prior to constructing the nearest-neighbor graph. Subsequently, $n$ random walks of length $w$ are initiated from each $x_\lambda$ (Figures~\ref{fig:2}{B} and~\ref{fig:2}{C}). These random walks are second-order Markov processes, wherein the probability $\rho_{bc}$ of transitioning from node $b$ to node $c$ depends on the previous node $a$ visited before $b$ according to the unnormalized transition probabilities
\begin{equation}
    \rho_{bc} =
\begin{cases}
1/\alpha, & \text{if } s_{ac} = 0 \\
1, &  \text{if } s_{ac} = 1 \\
1/\beta, &  \text{if } s_{ac} = 2
\end{cases},
\end{equation}
where $s_{ac}$ denotes the shortest path (minimum number of edges) connecting nodes $a$ (previously occupied node) and $c$ (subsequent node), while $\alpha > 0$ and $\beta > 0$ are parameters controlling the walker's bias. Smaller values of $\alpha$ favor more exploitative walks that sample local neighborhoods, while smaller values of $\beta$ favor more exploratory ones that sample deeper neighborhoods. 

After the walks are performed, we obtain $N \times n$ time series of length $w$, which we denote by $\{\tilde{x}_m\}_{m = 1,\dots, w}$. We then apply a symbolization process derived from the permutation entropy framework~\cite{bandt2002permutation} to each time series as follows. First, we select a sliding window of size $d$ (the embedding dimension) that traverses each $\{\tilde{x}_m\}$ (Figure~\ref{fig:2}{C}), dividing each trajectory into overlapping partitions $p_m$ represented by
\begin{equation}~\label{eq:partitions}
    p_m = (\tilde{x}_m, \tilde{x}_{m+1}, \dots, \tilde{x}_{m+d-1}),
\end{equation}
with $m = 1,\dots,w - d + 1$. For each partition $p_m$, we determine the permutation $\pi_m = (r_0, r_1,\dots, r_{d-1})$ of the index numbers $(0, 1, \dots, d - 1)$ that sorts the elements of $p_m$, such that $\tilde{x}_{m + r_0} \leq \tilde{x}_{m + r_1} \leq \dots \leq \tilde{x}_{m + r_{d-1}}$. In the event of equal values within a partition, that is, $x_{m+r_{k-1}} = x_{m+r_k}$, we impose $r_{k-1} < r_{k}$ for $k=1,\dots,d-1$~\cite{cao2004detecting}, thereby effectively treating earlier observations as smaller than later ones. As an illustration of the symbolization, consider a short walk $\{\tilde{x}_m\} = \{6,1,6,4,8\}$ with sliding window size $d = 3$. The first partition is $p_1 = (6,1,6)$ and sorting its elements yields $1 < 6 \leq 6$, that is, $x_{1+1} < x_{1+0} \leq x_{1+2}$. Thus, the permutation associated with $p_1$ is $\pi_1 = (1,0,2)$. It is important to note that the number of different permutation types each $\pi_m$ can assume is equal to $d!$. Accordingly, the complete set of all permutation types is represented by $\{\Pi_j\}_{j = 1, \dots, d!}$. For instance, if $d = 3$, there are $3! = 6$ permutations types: $\Pi_1 = (0,1,2)$, $\Pi_2 = (0,2,1)$, $\Pi_3 = (1, 0, 2)$, $\Pi_4 = (1, 2, 0)$, $\Pi_5 = (2, 0, 1)$, and $\Pi_{6} = (2,1,0)$.

By repeatedly applying the previously described symbolization method to all sampled walks and aggregating the corresponding permutations, we obtain a single symbolic sequence $\{\pi_m\}_{m = 1,\dots,M}$ comprising $M = N \times n \times (w-d+1)$ permutation symbols. The probability $\rho_{i}({\Pi_{i}})$ for each permutation type is then estimated as
\begin{equation}
    \rho_i(\Pi_i) = \frac{\text{number of permutations of type } \Pi_i \text{ in } \{\pi_m\}_{m = 1,\dots,M}}{N \times n \times (w-d+1)}.
\end{equation}
Finally, given the probability distribution $\mathcal{P} = \{\rho_i(\Pi_i)\}_{i=1,\dots,d!}$, the nearest neighbor permutation entropy~\cite{voltarelli2024characterizing} is computed via the normalized Shannon entropy (Figures~\ref{fig:2}{D} and~\ref{fig:2}{E}):
\begin{equation}\label{eq:knnnpe}
    H[\mathcal{P}] = -\frac{1}{\ln{d!}} \sum_{i = 1}^{d!} \rho_i(\Pi_i) \ln{\rho_i(\Pi_i)}.
\end{equation}
According to this equation, the $k$-nearest neighbor permutation entropy is restricted to $H \in [0,1]$, with $H$ quantifying the degree of irregularity in the spectrum $\{x_{\lambda}\}_{\lambda = \lambda_{\rm min}, \dots, \lambda_{\rm max}}$. In general, we expect $H\approx1$ when a single permutation dominates the probability distribution, whereas $H\approx0$ indicates a uniform distribution across all permutation types. The procedure for computing the $k$-nn permutation entropy is illustrated in Figure~\ref{fig:2}, and for the numerical implementation, we rely on the Python module \texttt{knnpe}~\cite{ribeiro2024knnpegithub}.

In applying the $k$-nearest neighbor permutation entropy to our absorbance spectra, we adopt the default values for the four parameters associated with the random walks~\cite{voltarelli2024characterizing}: $n = 10$, $w = 10$, $\alpha = 10$, and $\beta =0.001$. In contrast, the parameters $k$ and $d$ are extensively varied to identify optimal combinations. Additionally, to provide a basis for comparison, we also calculate the conventional permutation entropy of the spectra~\cite{bandt2002permutation}. Briefly, the permutation entropy is calculated by symbolizing a spectrum $\{x_\lambda\}_{\lambda = \lambda_{\rm min}, \dots, \lambda_{\rm max}}$ using a sliding window of size $d$ to partition the complete sequence of absorbance values. Next, we identify the permutations that sort these partitions (forming the permutation sequence $\{\pi_t\}$), determine the probabilities associated with each permutation type $[\rho_i (\Pi_i), i = 1, \dots, d!]$, and compute the corresponding Shannon entropy, implemented using the Python module \texttt{ordpy}~\cite{pessa2021ordpygithub}. Supplementary Table~S2 summarizes the parameters and key variables related to the $k$-nearest neighbor permutation entropy.

\section*{Results}

Leveraging the absorbance spectra collected during our depressurization trials, we investigate phase transitions in mixtures of carbon dioxide and a distilled petroleum fraction via an anomaly detection approach. Specifically, we calculate the $k$-nearest neighbor entropy for each spectrum recorded at a given pressure and monitor the dispersion of these entropy values as an indicator of phase transitions. 

Formally, we interpret a sequence of entropy values $\{H_P\}_{P = P_{\rm max}, \dots, P_{\rm min}}$ obtained from a depressurization trial as a time series indexed by decreasing pressure values, from $P_{\rm max}$ to $P_{\rm min}$. Thus, the trend and fluctuations in $H_P$ are quantified iteratively by calculating the average entropy
\begin{equation}
    \mu_P = \sum_{P' = P_{\rm max}}^{P} \frac{H_{P'}}{N}
\end{equation}
and its standard deviation 
\begin{equation}
    \sigma_P = \sqrt{\sum_{P' = P_{\rm max}}^{P} \frac{(H_{P'}-\mu_{P'})^2}{N}}
\end{equation}
evaluated over an expanding window indexed by pressures $P \in \{P_{\rm max}, \dots, P_{\rm min}\}$ as the depressurization proceeds. We assume that the transition pressure $P_t^{k{\rm nn}}$ associated with a phase transition is signaled by an abrupt change in $H_P$. Specifically, $P_t^{k{\rm nn}}$ is identified as the pressure at which $H_P$ first falls outside the confidence band defined by the thresholds $H^{-}_P = \mu_P - \gamma\sigma_P$ and $H^{+}_P = \mu_P + \gamma\sigma_P$, where $\gamma$ controls the band width. Supplementary Table~S3 summarizes the parameters and key variables used in our anomaly detection framework. 

\begin{figure*}[!ht]
  \centering
  \includegraphics[width=.85\textwidth]{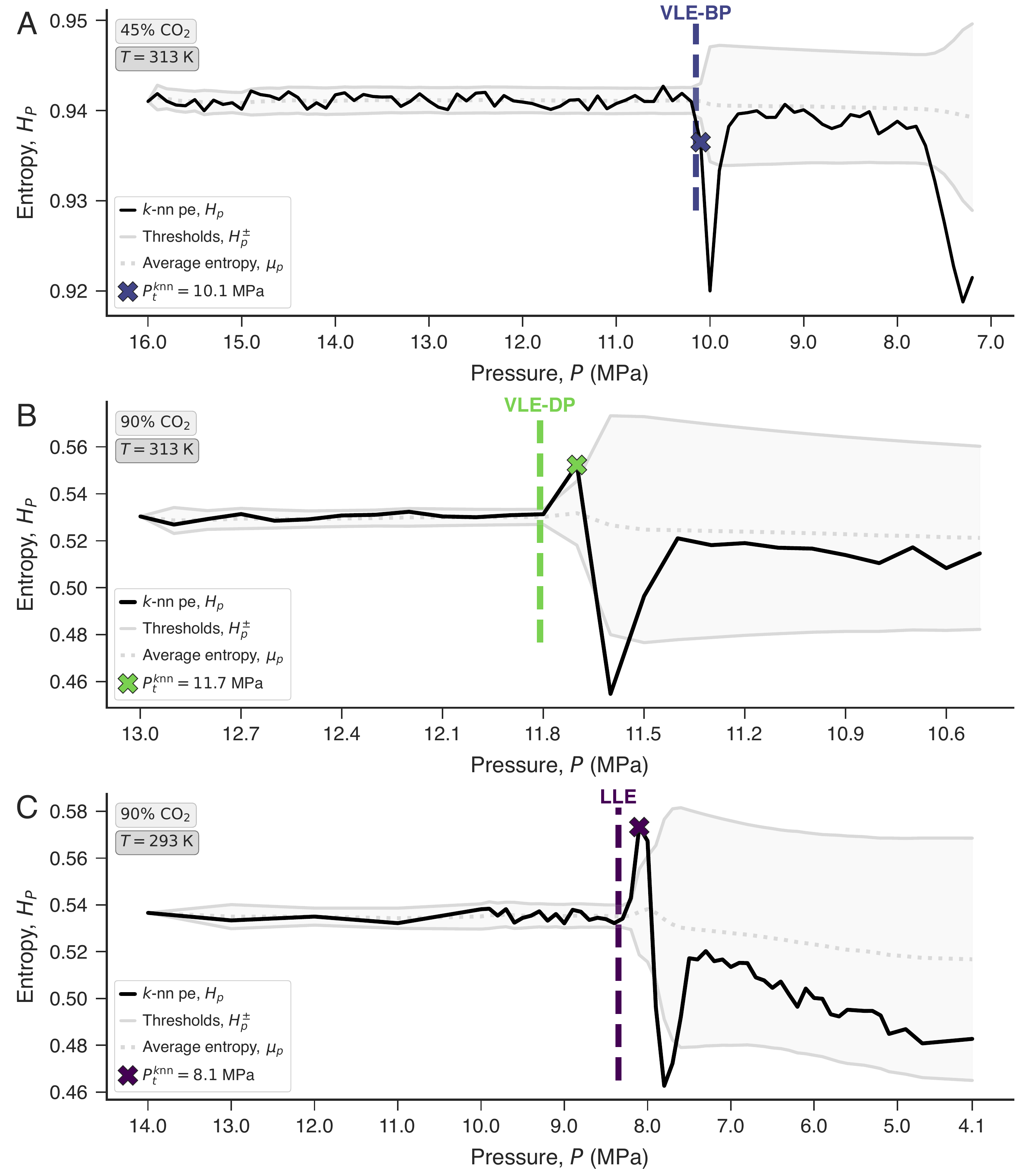}
  \caption{Phase transitions manifest as anomalies in entropy values. Panels (A), (B), and (C) show the values of $k$-nearest neighbor permutation entropy ($H_P$, solid black line) calculated from spectra of carbon dioxide and distilled petroleum fraction mixtures collected during depressurization trials. Light gray lines denote the confidence bands defined by threshold values $H_P^-$ and $H_P^+$, whose exceedance indicates a phase transition. Gray dotted lines display the average entropy values ($\mu_P$) calculated within expanding windows. Transition pressures visually determined by experts are indicated by colored vertical lines alongside the type of initiating phase, while colored cross markers show transition pressures identified by entropy values outside the confidence bands. ${\rm CO}_2$ concentration levels and temperatures are annotated within the panels. In all panels, the entropy parameters are set to $k = 265$ and $d = 5$, with the threshold parameter $\gamma = 2.25$. 
  }
\label{fig:3}
\end{figure*}

Figure~\ref{fig:3} illustrates this procedure for three different samples. For instance, Figure~\ref{fig:3}{A} presents a single depressurization trial involving a mixture containing  $45\%$ ${\rm CO}_2$ at $T = 313$ K. The initially homogeneous liquid mixture transitions to a vapor-liquid equilibrium upon crossing the bubble-point curve (VLE-BP) at $P_t^{\rm exp} = 10.2$~MPa. Notably, the spectra entropy values first exceed the upper threshold $H_P^+$ at $P_t^{k{\rm nn}} = 10.1$~MPa, signaling the onset of the VLE-BP phase in close agreement with the visually inferred transition pressure. Figure~\ref{fig:3}{B} shows a transition from an initially homogeneous gas mixture to vapor-liquid equilibrium by crossing the dew-point curve (VLE-DP), while Figure~\ref{fig:3}{B} depicts a transition from an initially homogeneous liquid mixture to a two-phase liquid-liquid equilibrium (LLE). For the VLE-DP transition in Figure~\ref{fig:3}{B}, our method identifies $P_t^{k{\rm nn}} = 11.7$ MPa, marginally smaller than the visually determined $P_t^{\rm exp} = 11.8$ MPa, and for the LLE transition in Figure~\ref{fig:3}{C} it predicts $P_t^{k{\rm nn}} = 8.1$ MPa, slightly lower than the visually determined $P_t^{\rm exp} = 8.4$ MPa.

We next apply our approach to characterize phase transitions across all 42 depressurization trials. As the approach depends on the threshold parameter $\gamma$ and the entropy parameters $k$ and $d$, we systematically explore all possible combinations of $\gamma \in \{1.00, 1.05, \dots, 3.50\}$, $k \in \{10,15,\dots,300\}$, and $d \in \{3,4,5\}$ to optimize the predictive performance and maximize the number of accurately detected transitions. To identify the best-performing parameter set, we adopt a leave-one-out cross-validation strategy~\cite{james2023introduction} in which each depressurization trial is iteratively used as a validation set, while the remaining trials constitute the training set. This procedure ensures that the parameter optimization process remains strictly independent of the evaluation data, thereby enabling out-of-sample predictions that reflect our method's generalizability and predictive power. Through this extensive search, we find that the best results are consistently obtained for $\gamma = 2.25$, $k = 265$, and $d = 5$ for 41 out of the 42 validation trials. The only exception is a trial prepared with ${\rm CO}_2 = 45\%$ and $T = 333$ K, where no phase transition was detected and a slightly different parameter set emerged. 

\begin{figure*}[!ht]
  \centering
  \includegraphics[width=1\textwidth]{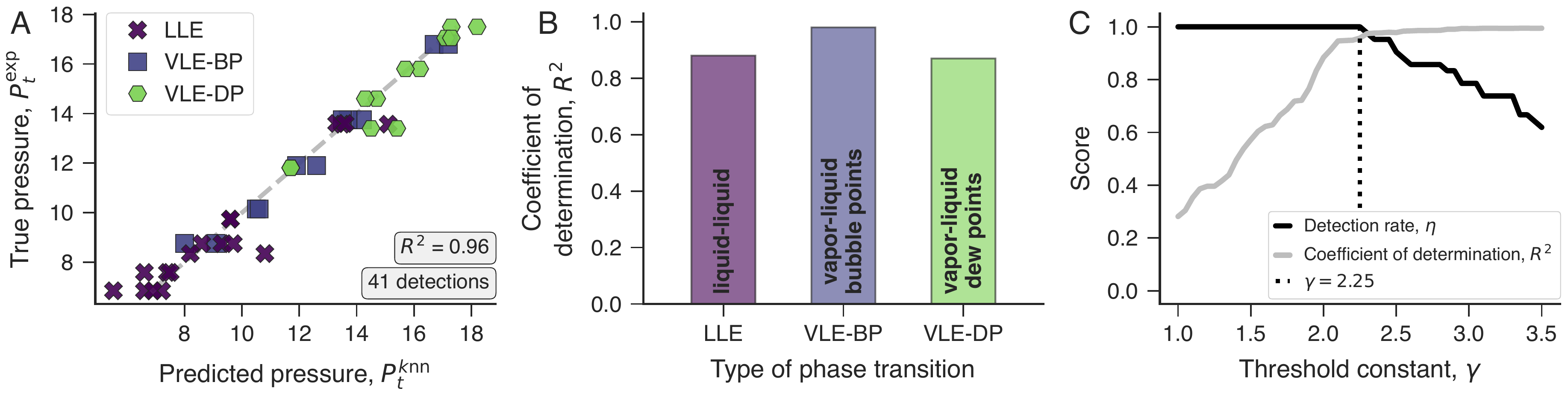}
  \caption{Accuracy of out-of-sample predictions for transition pressures in mixtures of carbon dioxide with a distilled petroleum fraction. (A) Relationship between the true $P_t^{\rm exp}$ (experimentally determined) and predicted $P_t^{\rm k{\rm nn}}$ transition pressures for 41 of 42 depressurization trials in which a transition was successfully identified. Colors indicate the transition types observed during the depressurization of homogeneous mixtures. The coefficient of determination $(R^2)$ quantifying the agreement between true and predicted pressures is shown within the panel. (B) Coefficient of determination $(R^2)$ stratified by the transition types. Predictions were obtained using a leave-one-out cross-validation approach, with the threshold parameter $\gamma$ and entropy parameters $k$ and $d$ optimized independently of the evaluation data. (C) Fraction of detected transitions $\eta$ (black line) and coefficient of determination $R^2$ (gray line) as functions of the threshold parameter $\gamma$, with entropy parameters fixed at their optimal values ($k = 265$ and $d = 5$).}
\label{fig:4}
\end{figure*}

Figure~\ref{fig:4}A compares the predicted transition pressures ($P_t^{\rm knn}$) with the visually determined reference values ($P_t^{\rm exp}$) for the 41 trials in which a phase transition was detected (see also Supplementary Table~S4). The predictions closely match the experimental values, with a coefficient of determination $R^2 = 0.96$ and a mean absolute percentage error (MAPE) of $4.51\%$. Figure~\ref{fig:4}B further shows that this high predictive performance holds across different types of initiating phase equilibria. For example, transitions to LLE yield $R^2 = 0.88$ and ${\rm MAPE} = 6.25\%$, transitions to VLE-BP lead to $R^2 = 0.98$ and ${\rm MAPE}  = 3.16\%$, and transitions to VLE-DP produce $R^2 = 0.87$ and ${\rm MAPE} = 3.15\%$. In 22 out of 41 detected transitions, the predicted transition pressures are equal to or smaller than the true values, indicating an approximately symmetric chance of anticipating or lagging the experimentally determined transitions. On average, the absolute difference between the predicted pressures and the pressures sampled nearest to the experimental estimates is $0.5$ MPa. These discrepancies may partly arise from the finite temporal resolution of the experimental measurements and the windowed nature of the $k$-nn permutation entropy calculation, both of which can hinder the detection of abrupt changes; gradual variations in system dynamics near the transition may also contribute. Additionally, given the inherent difficulty in exhaustively monitoring these rapid transitions, visually determined transition pressures are themselves susceptible to experimental errors.

Furthermore, given the stability of the optimal entropy parameters across validation trials, we investigate the effect of the threshold parameter $\gamma$ on predictive performance while fixing $k = 265$ and $d = 5$ -- their optimal values. As shown in Figure~\ref{fig:4}{C}, values of $\gamma$ in the range $1 \leqslant \gamma \leqslant 2.25$ yield the highest detection rate $\eta$, albeit at the cost of reduced accuracy in the predicted transition pressures. Conversely, for $2.00 \leqslant \gamma \leqslant 3.50$, the predictions of transition pressures improve, but the fraction of detected transitions decreases with increasing $\gamma$. Since $\gamma$ controls the width of the confidence band, these findings indicate that transition pressures are more accurately identified when entropy changes are more abrupt. Thus, enhancing the temporal resolution of the experimental measurements would likely improve both detection accuracy and performance, as the confidence band would become narrower and smoother.

For further comparison, we apply our anomaly detection approach using the standard permutation entropy in place of the $k$-nearest neighbor entropy. As before, we employ a leave-one-out cross-validation strategy to identify the optimal parameter combination, considering $\gamma \in {1.00, 1.05, \dots, 3.50}$ and $d \in {3, 4, 5}$. This modification, however, results in markedly poorer predictive performance, with $R^2 = 0.67$ and ${\rm MAPE} = 15.0\%$, despite correctly detecting all 42 transitions (Supplementary Figure~S1A). Stratifying results by transition type further reveals that standard permutation entropy performs particularly poorly for LLE and VLE-BP transitions (Supplementary Figure~S1B). These findings indicate that the superior performance of our approach stems from the $k$-nn permutation entropy's ability to exploit local neighborhood relationships via the graph-based sampling strategy that captures amplitude variations and wavelength gaps ignored by the standard permutation entropy, thereby enhancing noise resilience and increasing sensitivity in detecting phase transitions.

\begin{figure*}[!ht]
  \centering
  \includegraphics[width=0.75\textwidth]{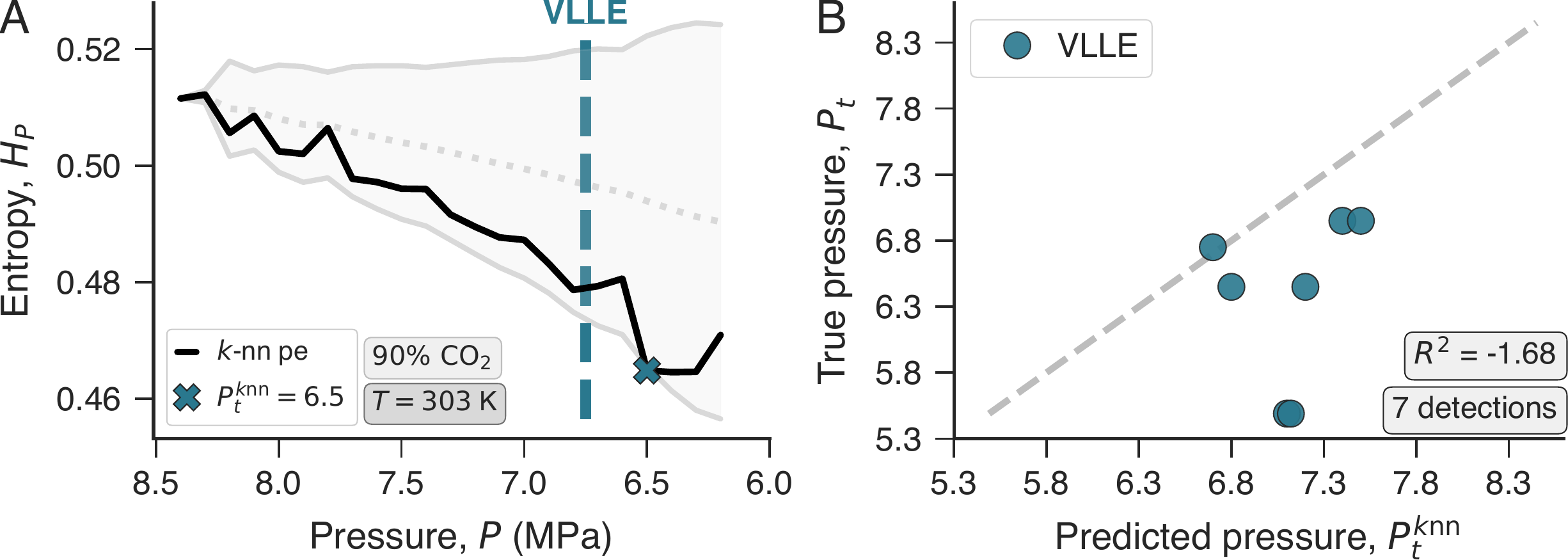}
  \caption{Challenges in predicting transition pressures between liquid–liquid and vapor–liquid–liquid equilibria. (A) Values of the $k$-nearest neighbor permutation entropy ($H_P$, solid black line) calculated from spectra of carbon dioxide and distilled petroleum fraction mixtures during depressurization trials involving a transition from a liquid-liquid equilibrium to vapor-liquid-liquid equilibrium. Light gray lines represent threshold values whose exceedance indicates a phase transition ($H_P^-$ and $H_P^+$), while gray dotted lines show average entropy values ($\mu_P$). The experimentally identified transition pressure ($P_t^{\rm exp}$) is marked by a vertical colored line, and the predicted transition pressure ($P_t^{k{\rm nn}}$) is shown by a colored cross. Mixture composition and temperature are indicated within the panel. (B) Relationship between the true $P_t^{\rm exp}$ (experimentally determined) and predicted $P_t^{k\rm nn}$ transition pressures for 7 of 9 depressurization trials exhibiting a second phase transition. The coefficient of determination $(R^2)$ quantifying the agreement between true and predicted pressures is shown within the panel.
}
\label{fig:5}
\end{figure*}

Continuing our analysis, we observe that nine depressurization trials exhibit a second transition, each corresponding to a change from liquid-liquid equilibrium (at higher pressures) to vapor-liquid-liquid equilibrium (at lower pressures). Figure~\ref{fig:5}{A} illustrates the evolution of the $k$-nearest neighbor permutation entropy in one such trial, along with the confidence band and the detected transition obtained using our approach with previously optimized parameters ($\gamma=2.25$, $k =265$ and $d = 5$). However, unlike this particular example, and despite detecting all transitions, our method generally performs poorly in identifying this class of transitions, yielding a negative coefficient of determination ($R^2 = -2.42$) and a substantial prediction error (mean absolute percentage error, ${\rm MAPE} = 16.1\%$). In an attempt to address this limitation, we re-optimize the parameter set using a leave-one-out cross-validation strategy tailored to this class of transitions. Yet, as shown in Figure~\ref{fig:5}B (see also Supplementary Table~S5), even re-optimized parameters fail to significantly improve performance, with the coefficient of determination remaining negative ($R^2 = -1.68$) and a high prediction error persisting ($\mathrm{MAPE} = 13.0\%$), based on 7 detected transitions out of the 9 observed. Furthermore, replacing $k$-nearest neighbor entropy with standard permutation entropy results in even less accurate predictions. These results indicate that transitions between liquid-liquid equilibrium (LLE) and vapor-liquid-liquid equilibrium (VLLE) are considerably more difficult to detect via abrupt changes in entropy than transitions between homogeneous mixtures and other types of high-pressure phase equilibria. This increased difficulty likely stems from the greater complexity of the LLE–VLLE transition -- characterized by vapor nucleation and mass transfer between phases -- which renders phase homogeneity more difficult to achieve experimentally~\cite{borges2015near}. Additionally, due to the spatial arrangement of phases within the optical cell, the NIR beam may predominantly sample only one of the coexisting phases -- typically the denser or more spectrally active liquid -- while underrepresenting the emerging vapor phase or the less optically absorbing liquid. The relatively weak absorbance of the vapor phase, combined with its limited residence time or distribution along the optical path, may further diminish its contribution to the spectra. As the $k$-nn entropy reflects spectral complexity rather than intensity alone, the absence of abrupt or structured spectral changes across scans makes such transitions inherently more difficult to resolve. Finally, because these transitions occur later in the depressurization process, the number of spectra preceding the transition is limited, reducing the temporal resolution and the robustness of confidence band estimation -- further constraining the effectiveness of our detection approach.

\section*{Discussion and Conclusions}\label{sec:conclusions}

We have presented an investigation of phase transitions in synthesized mixtures of carbon dioxide with a distilled petroleum fraction using an anomaly detection methodology. During depressurization trials, our approach repeatedly calculates the $k$-nearest neighbor permutation entropy -- an information-theoretic measure directly computed from near-infrared absorbance spectra -- and signals the occurrence of a phase transition by identifying large variations in entropy values. In line with previous studies on high‐pressure systems~\cite{dohrn2024high-pressure, dohrn2012experimental}, our work integrates synthetic and analytical methods to elucidate phase behavior in high-pressure conditions, employing a modified experimental setup that enables continuous system monitoring without requiring sample extraction~\cite{borges2015near}. 

Our methodology yields robust results, accurately identifying transitions from initially homogeneous mixtures to different types of two-phase equilibria. Moreover, the approach can be tailored to specific mixtures of carbon dioxide and hydrocarbon fractions (for instance, $60\%$ ${\rm CO}_2$) or particular types of phase transitions, allowing for further improvements in performance. Indeed, detecting phase transitions in high-pressure hydrocarbon mixtures with elevated ${\rm CO}_2$ content has proven challenging, typically requiring cumbersome data preprocessing, specialized detection techniques based on field expertise, and qualitative visual assessments~\cite{borges2015near}, difficulties that our method overcomes through an online, streamlined, and computationally efficient strategy. 

Although we have used the same dataset as Borges \textit{et al.}~\cite{borges2015near}, the results are not directly comparable. In their study, transition pressures were estimated via visual identification of inflection points in cumulative absorbance curves, whereas our fully automated approach employs a leave-one-out cross-validation scheme to enable out-of-sample predictions. Moreover, their analysis reflects average behaviors across isothermal depressurization trials at fixed ${\rm CO}_2$ concentrations, while our method produces individualized predictions for each trial. We have further verified that the predictive power of our approach arises from the ability of $k$-nearest neighbor permutation entropy to exploit local neighborhood structure through a graph-based sampling strategy that captures both amplitude variations and wavelength gaps -- features not adequately resolved by standard permutation entropy, which yielded markedly inferior results. While future studies may explore the performance of other generalized variants of permutation entropy, we emphasize that the capacity to simultaneously account for gaps, amplitude values, and non-sequential sampling remains a distinctive feature of the $k$-nearest neighbor permutation entropy~\cite{voltarelli2024characterizing}.

Transitions between liquid-liquid and vapor-liquid-liquid equilibria, however, were not effectively captured by changes in the $k$-nearest neighbor permutation entropy. A closer examination reveals that, in these cases, entropy values tend to shift gradually, limiting our method's ability to detect abrupt changes. This observation represents an intriguing avenue for future studies aimed at defining or applying more sensitive information‐theoretic quantifiers capable of detecting gradual shifts. Beyond the intrinsic complexity of the liquid-liquid to vapor-liquid-liquid equilibria transition, part of the challenge may stem from the fact that the mixtures were not allowed to stabilize after reaching liquid-liquid equilibrium; consequently, the experimental conditions preceding this phase transition differ from those preceding the transition of the initially homogeneous mixture into liquid-liquid equilibrium. These transitions also occur later in the depressurization process, resulting in fewer spectra preceding their onset and limiting the reliability of confidence band estimation -- factors that further hinder the application of our approach in these cases. We believe that experiments specifically designed to probe such transitions, incorporating slower depressurization rates and using higher sampling rates for collecting the spectra, may yield more conclusive results. We also note that our anomaly-detection framework is somewhat computationally intensive due to the exhaustive search over a three-dimensional parameter space, which may become cumbersome for large spectral datasets.

Despite these limitations, our approach is sufficiently general to be adapted for studying phase behavior in other chemical systems subjected to high-pressure conditions and monitored by spectrophotometry. Thus, our work underscores the noninvasive nature of employing permutation entropy and ordinal methods in the investigation of complex physicochemical systems~\cite{sigaki2019estimating, pessa2021determining}.

\section*{Data availability}
All code and data necessary to reproduce the results and figures in this work are available at the GitLab repository \url{https://gitlab.com/complexlab/nir-co2oil-knnpe}.

\bibliography{references}

\section*{Acknowledgements}
The authors acknowledge the support of the Coordena\c{c}\~ao de Aperfei\c{c}oamento de Pessoal de N\'ivel Superior (CAPES), the Conselho Nacional de Desenvolvimento Cient\'ifico e Tecnol\'ogico (CNPq -- Grant 303533/2021-8), and Slovenian Research and Innovation Agency (Javna agencija za znanstvenoraziskovalno in inovacijsko dejavnost Republike Slovenije) (Grants P1-0403 and N1-0232).

\section*{Author contributions}
A.A.B.P., L.G.J.M.V., L.C.F., A.G.M.T., C.D., P.M.N., M.P., and H.V.R. designed research, performed research, analyzed data, and wrote the paper.

\clearpage


\section*{Supplementary material}

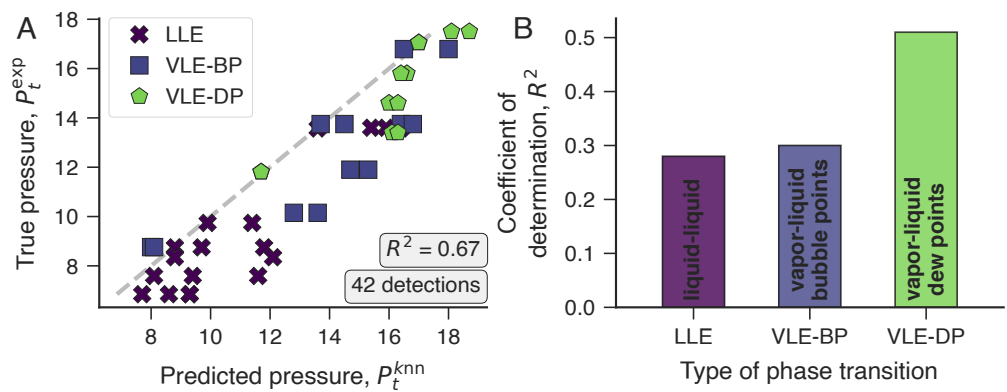

**Figure S1.** Accuracy of out-of-sample predictions for transition pressures using standard permutation entropy. (A) Relationship between the true $P_t^{\text{exp}}$ (experimentally determined) and predicted $P_t^{\text{pe}}$ transition pressures obtained by replacing the *k*-nearest neighbor permutation entropy with the standard permutation entropy in our anomaly detection framework. Data correspond to all 42 depressurization trials in which a phase transition was successfully identified. Colors indicate the transition types observed during the depressurization of homogeneous mixtures. The coefficient of determination ($R^2$) quantifying the agreement between true and predicted pressures is shown in the panel. (B) Coefficient of determination ($R^2$) stratified by the transition types. Predictions were obtained using a leave-one-out cross-validation approach, with the threshold parameter $\gamma$ and entropy parameter *d* optimized independently of the evaluation data. Results are markedly inferior to those reported in the main text using *k*-nearest neighbor permutation entropy.

| Variable | Description | Units |
|---|---|---|
| $\lambda$ | Absorbance | arbitrary units (a. u.) |
| $x_\lambda$ | Wavelength | nanometers (nm) |
| $P$ | Pressure | megapascal (MPa) |
| $T$ | Temperature | kelvin (K) |
| $P_t^{\exp}$ | Transition pressure visually determined | megapascal (MPa) |
| $P_t^{knn}$ | Transition pressure predicted by anomaly detection | megapascal (MPa) |

**Table S1.** Description of key variables associated with the spectrophotometric data obtained during depressurization trials of mixtures of carbon dioxide and distilled petroleum fractions.

| Variable | Description |
|---|---|
| $d$ | Size of sliding window (embedding dimension) |
| $k$ | Number of neighbors in nearest neighbor's graph |
| $w$ | Length of random walks in nearest neighbor's graph |
| $n$ | Number of random walks starting from each node in nearest neighbor's graph |
| $\beta$ | Exploration bias in random walks in nearest neighbor's graph |
| $\alpha$ | Exploitation bias in random walks in nearest neighbor's graph |
| $\rho(\Pi)$ | Probability of occurrence of permutation type $\Pi$ in the series sampled with the walks |
| $H$ | $k$-nearest neighbor permutation entropy |

**Table S2.** Description of key variables associated with the $k$-nn permutation entropy.

| Variable | Description |
|---|---|
| $\mu_P$ | Average entropy for spectra sampled between $P_{\max}$ and $P$ |
| $\sigma_P$ | Standard deviation of entropy for spectra sampled between $P_{\max}$ and $P$ |
| $\gamma$ | Threshold parameter |
| $H^+$ | Upper threshold for nonanomalous values of entropy |
| $H^-$ | Lower threshold for nonanomalous values of entropy |

**Table S3.** Description of key variables associated with the anomaly detection approach.

| Trial | CO$_2$ | Temperature (K) | True pressure (MPa) | Predicted pressure (MPa) |
|---|---|---|---|---|
| 1  | 35% | 293 | 6.85  | 7.20  |
| 2  | 35% | 293 | 6.85  | 6.60  |
| 3  | 35% | 293 | 6.85  | 5.52  |
| 4  | 35% | 293 | 6.85  | 6.90  |
| 5  | 35% | 303 | 7.59  | 6.60  |
| 6  | 35% | 303 | 7.59  | 7.40  |
| 7  | 35% | 303 | 7.59  | 7.50  |
| 8  | 35% | 313 | 8.76  | 9.10  |
| 9  | 35% | 313 | 8.76  | 8.00  |
| 10 | 35% | 333 | 11.90 | 12.60 |
| 11 | 35% | 333 | 11.90 | 11.90 |
| 12 | 45% | 293 | 8.75  | 9.30  |
| 13 | 45% | 293 | 8.75  | 8.60  |
| 14 | 45% | 293 | 8.75  | 9.70  |
| 15 | 45% | 313 | 10.15 | 10.50 |
| 16 | 45% | 313 | 10.15 | 10.60 |
| 17 | 45% | 333 | 13.75 | 14.20 |
| 18 | 45% | 333 | 13.75 | -.-   |
| 19 | 60% | 303 | 13.75 | 13.70 |
| 20 | 60% | 303 | 13.75 | 13.50 |
| 21 | 60% | 313 | 16.79 | 16.70 |
| 22 | 60% | 313 | 16.79 | 17.20 |
| 23 | 60% | 333 | 13.57 | 13.50 |
| 24 | 60% | 333 | 13.57 | 15.10 |
| 25 | 70% | 303 | 13.60 | 13.30 |
| 26 | 70% | 303 | 13.60 | 13.60 |
| 27 | 70% | 313 | 14.60 | 14.70 |
| 28 | 70% | 313 | 14.60 | 14.30 |
| 29 | 70% | 333 | 17.50 | 17.30 |
| 30 | 70% | 333 | 17.50 | 18.20 |
| 31 | 80% | 313 | 13.40 | 15.40 |
| 32 | 80% | 313 | 13.40 | 14.50 |
| 33 | 80% | 333 | 17.05 | 17.10 |
| 34 | 80% | 333 | 17.05 | 17.30 |
| 35 | 90% | 293 | 8.35  | 1.0.0 |
| 36 | 90% | 293 | 8.35  | 8.20  |
| 37 | 90% | 303 | 9.74  | 9.60  |
| 38 | 90% | 303 | 9.74  | 9.60  |
| 39 | 90% | 313 | 11.81 | 11.70 |
| 40 | 90% | 313 | 11.81 | 11.70 |
| 41 | 90% | 333 | 15.80 | 16.20 |
| 42 | 90% | 333 | 15.80 | 15.70 |

**Table S4.** Carbon dioxide (CO$_2$) concentration levels and temperatures relative to 6 mixtures submitted to 42 depressurization trials contained in our dataset. Experimentally determined transition pressures (True pressure) and transition pressures determined by the anomaly detection (Predicted pressure) are also shown inside the table.

| Trial | CO$_2$ | Temperature (K) | True pressure (MPa) | Predicted pressure (MPa) |
|---|---|---|---|---|
| 1 | 45% | 293 | 5.49 | 7.10 |
| 2 | 45% | 293 | 5.49 | 7.10 |
| 3 | 45% | 293 | 5.49 | – |
| 4 | 60% | 303 | 6.95 | 7.40 |
| 5 | 60% | 303 | 6.95 | 7.50 |
| 6 | 70% | 303 | 6.45 | 7.20 |
| 7 | 70% | 303 | 6.45 | 6.80 |
| 8 | 90% | 303 | 6.75 | – |
| 9 | 90% | 303 | 6.75 | 6.70 |

**Table S5.** Carbon dioxide (CO$_2$) concentration levels and temperatures relative to 4 mixtures submitted to 9 depressurization trials contained in our dataset that present transitions between liquid-liquid-equilibria and a vapor-liquid-liquid equilibria. Experimentally determined transition pressures (True pressure) and transition pressures determined by the anomaly detection (Predicted pressure) are also shown inside the table.


\end{document}